\parindent=0pt
\parskip=20pt
\magnification=1200

\rightline {KCL-TH-95-9}
\rightline{hep-th/9509140}
\vskip 1in
\centerline { {\bf{  Non-Perturbative Green's Functions in}}}
\centerline{{\bf Theories with Extended Superconformal Symmetry}}
\vskip 1in
{\centerline {\bf{ Paul Howe and Peter West} }}
\par
{\centerline {Mathematics Department}}
\par
{\centerline {King's College}}
\par
{\centerline {Strand}}
\par
{\centerline {London WC2R 2LS}}
\vskip 1in
{\underbar {Abstract}}
\par

The multiplets that occur in four dimensional rigidly supersymmetric theories
can
be  described either by chiral superfields in Minkowski superspace or analytic
superfields in harmonic superspace. The superconformal Ward identities for
Green's functions of gauge invariant operators of these types are derived. It
is shown that there are no chiral superconformal invariants.
It is further shown
that   the Green's functions of analytic operators are
severely restricted  by the superconformal Ward when analyticity is
taken into account.

\vfil
\eject

\parskip=18pt
The most symmetric
four dimensional quantum field theory with particles of spin
less than or equal to one  is the
$N=4$ supersymmetric Yang-Mills theory [1] with four rigid
supersymmetries. The action for this  theory is uniquely determined by its
symmetries once we specify the gauge group.  Theories which possess
two rigid supersymmetries and have spin less than or equal to one can
only be formed from  a coupling of
$N=2$ Yang-Mills [2] to
$N=2$ matter [3] These theories are also uniquely determined by their
symmetries once we specify their gauge group  and
 the representation of the $N=
2$ matter. In effect, these are the most
symmetric rigidly supersymmetric theories, since if   we include
particles of spin 3/2 and above then we can only have  causal
propagation if we also include gravity and so have a theory of local
supersymmetry.  The special nature of the rigidly  extended
supersymmetric theories
 was further confirmed when it was found that the $N=4$ Yang-Mills
theory is  finite or superconformally invariant [4] and that the  $N=2$
theories
have only a  one loop beta function [20,21,5,22]. For
a large class of these latter theories it was found [5] that this one
loop beta function
 vanishes so these theories are also   superconformally invariant.
More recently [6], by  exploiting  duality, a non-perturbative expression
for a certain sector of the effective action of  $N=2$ Yang-Mills with  $SU(2)$
gauge group has been deduced.  The duality conjecture has also been extended
[7] to the simplest finite $N=2$ theory with $SU(2)$
gauge group.
\par
Since the extended rigidly supersymmetric  theories are essentially
 determined by their symmetries one might hope that one
could determine their gauge invariant Green's functions
solely by using these symmetries. For the
finite theories this symmetry is an extended  superconformal
supersymmetry. However, it is well known [8] that although (super)
conformal symmetries  determine the form of two and three point
Green's functions they  do not, in general,
determine the higher point functions.
 A spectacular  exception [9] to this rule
is provided by a class of conformally invariant two dimensional models called
minimal models. These  can be solved as a consequence of the occurrence
of null states in  the representations of the conformal group which
they carry. The elimination of these null states
leads to  differential equations for the Green's functions which can
then be determined.
\par
One might hope that if similar techniques could be used in four
dimensional theories then the best candidates would be the most
symmetric  conformally invariant theories, i.e. theories with extended rigid
supersymmetry. In four dimensions the (super) conformal
group consists of  only the generalisation of those transformations which
are globally defined in two dimensions. As is to be expected these can
be used to  find explicitly the two and three point Green's functions.
One indication that one may be able to solve for higher point Green's
functions came from the study of the N=2 supersymmetric minimal models
in two dimensions. These models contain certain primary fields which
correspond to chiral superfields and it was   shown [10]
that  the Green's functions  of chiral superfields of the same
chirality could be found explicitly  using only the globally defined
two dimensional superconformal transformations and the chiral
constraint which is equivalent to  the elimination of one of the
null states. This  result leads one to hope that one can solve for
all the Green's functions which are for chiral superfields  of the
same chirality in any supersymmetric theory at a fixed point.
Following the earlier work of reference [12], one step in this direction, was
taken in reference [11] where it was
shown that for a supersymmetric field theory at a fixed point the
anomalous dimensions of gauge invariant chiral superfields were
determined by their R weight. This result was used [11] to argue for
the triviality of the Wess-Zumino model and super Q.E.D.
\par
In this paper, we shall argue that  superconfromal symmetry
places very strong
constriants on Green's functions in any chiral or
analytic sector of a superconformal theory.  By a chiral Green's
function   we mean one that involves only all chiral or anti-chiral
superfields. Such sectors  can found in N=1 and N=2 rigidly
supersymmetric theories. In the former they involve the
$N=1$ Wess-Zumino multiplet and the
$N=1$ Yang-Mills field strength, while in the latter they
involve the $N=2$ Yang-Mills field strength. Analytic
Green's functions are similarly defined with analytic rather
than chiral superfields. Such superfields occur in N=2 and N=4
rigidly supersymmetric theories and are constructed using
harmonic superspace [13] as will be explained below.
Specifically, analytic superfields can be used to describe  the $N=2$
hypermultiplet [13] and the
$N=4$ Yang-Mills field strength multiplet [14]. An important feature of
analytic
superfields is that their conjugates are of the same analytic type  in
contrast to the chiral case. Hence, all multiplets  of rigidly
supersymmetric theories can be described by chiral or analytic
superfields.   We shall derive  the superconformal  Ward identities
for correlators involving such superfields and show that there are no
chiral invariants and that the form of the four point analytic superconformal
invariants is severely restricted.
\par
The chiral and analytic constraints are an important feature of superspace
formulations of
supersymmetric theories. We recall that
 a  chiral superfield satisfies the constraint $D_{\dot
\alpha } ^i\phi =0,\ \dot \alpha=1,2, \ \ i=1, \ldots ,N$
 where $\phi$ is the superfield in question, $N$  the number of
supersymmetries and $D_{\alpha }^i \equiv {\partial \over \partial
\theta^{\alpha i}}-
{i\over 2} \theta^{\dot \alpha} _i{\partial \over
\partial x^{\alpha \dot \alpha} }$ ,
$D_{\dot \alpha }^i \equiv -{\partial \over \partial
\theta^{\dot \alpha }_i}+
{i\over 2} \theta^{\ \alpha i}{\partial \over
\partial x^{\alpha \dot \alpha} }$
are the Minkowski
 superspace covariant derivatives. The chiral superfields
which occur in rigidly supersymetric theories are  the $N=1$
Wess-Zumino multiplet $\varphi$,    the $N=1$  Yang-Mills field
strength superfield $W_\alpha$ and the
 $N=2$  Yang-Mills field strength superfield
  $W$. In fact, these field
strengths obey chirality constraints with gauge covariant
superspace derivatives, but gauge invariant products of these
field strengths obey ordinary chiral constraints.  The above
superfields are defined on Minkowski superspace which
 has coordinates
$\{ x^{\alpha
\dot \alpha},\  \theta^{\alpha i},\   \theta^{\dot \alpha }_i\}$, $i=1,\ldots
N$.
 As a result of the chiral constraint a chiral superfield $\phi$
depends  only on
$s^{\alpha
\dot \alpha}= x^{\alpha
\dot \alpha} - {i\over2}  \theta^{\alpha j} \theta^{\dot \alpha}_ j$
and $\theta^{\alpha i}$ and  can be regarded as a function on
chiral superspace. While Minkowski superspace  is  the coset space of
the super Poincar\'e group divided by the Lorentz group, chiral
superspace   can also be interpreted  as a coset space which is obtained
from the  complexified super Poincar\'e group by
dividing by   a suitable sub-supergroup which is generated by  the
Lorentz group and a chiral supersymmetry. Chiral constraints on superfields
have  important consequences for quantum corrections due to  the well known
non-renormalization
theorem [15]. This  states that all corrections to the effective action arise
as  integrals over the whole of superspace and not over a chiral subspace. In
fact,  if massless particles are present,  the chiral superpotential
generically receives quantum corrections [16] ( so called non-holomorphicity ),
the first  example of which was found in reference [10].
\par
The other  multiplets of rigid supersymmetry , namely
$N=2$ matter and N=4 Yang-Mills are also described by
constrained superfields.  The
$N=2$ matter  multiplet (the hypermultiplet) was originally described
[3] by a scalar superfield
$\phi_i,\ i=1,2$ which  transforms as a complex doublet of $SU(2)$ and
satisfies the superspace constraints
$$D_{\alpha (i} \phi_{j)}=0
=D_{\dot \alpha (i}\phi_{j)}
\eqno (1)$$
where the internal  indices are raised and lowered
with the $\epsilon$ tensor, for example,
$D_{\dot \alpha i} =  D_{\dot \alpha}^j \epsilon_{j i}$.
 The $N=4$  Yang-Mills field strength superfield is a Lorentz scalar
superfield
$W_{ij}$ which transforms under the six dimensional representation
of $SU(4)$, and as such  is anti-symmetric,
$W_{ij} = -W_{j i} $ and self-dual $W_{ij}= {1\over 2}\epsilon
_{ijkl}\bar W^{kl}$ . It satisfies the constraint
$$D_{\alpha i}W_{jk}= D_{\alpha [i}W_{jk]}.
\eqno(2)$$
 In contrast to the chiral constraints,
both the above superfields are on-shell and there appear to be
no associated analogues of chiral superspace. Using  harmonic
superspace [13,14], we now discuss the alternative formulations of
these multiplets which overcome this drawback.
\par
Harmonic superspace extends the usual
Minkowski superspace by an internal coset space. The $N=2$
harmonic superspace is $\hat M_2 = M_2\times S^2$ where $M_N$
denotes $N$-extended Minkowski superspace. The additional space
$S^2$ can best be thought of as
${SU(2)\over U(1)}$\footnote*{Throughout this paper all coset spaces are spaces
of right cosets.}. For $N=4$, the harmonic space of interest to us
here is $\hat M_4 = M_4\times Gr_2(4)$ where $Gr_2(4)$ is the
Grassmannian of two planes in ${\bf {C}}^4$. It can be described as the
coset space $SU(4) \over S(U(2)\times U(2))$.  One could equally
well consider $\hat M_N$ as  a coset space of the super Poincar\'e
group with the  isotropy group taken to be the product of the Lorentz group and
one of the above
internal isotropy groups. Superfields  on
$\hat M_N$ are taken to carry induced representations of $G=SU(N)$ with
the isotropy group $H$  corresponding to those of the above cosets.
Such representations act on  fields $\varphi$ which
carry  a matrix representation of $H$ denoted by $D(h)$ for
any $h\in H$. The fields $\varphi$  can be regarded as being defined
on either $M_N\times G$ or $M_N\times {G\over H}$. In the former case, the
field $\varphi$ is taken to be equivariant with respect to the left
action of the isotropy group, meaning it satisfies  the relation
$\varphi(hg) = D(h) \varphi (g),$ $ \forall\  h \ \in\  H$. The
induced representation is then defined to act as
$U(g_1)\varphi(g) =\varphi ^\prime(g) = \phi(gg_1) \ ,\
\forall g,g_1\ \in\  G$. The coset space description can be recovered
by choosing
 a local section of $G$ considered as a left principal $H$-bundle
over ${G\over H}$ or, more plainly, by  choosing coset representatives. We
denote by $y$ the corresponding coordinates on ${G\over H}$ inherited from
those of the  group.  We define the field $\phi$ on ${G\over H}$ as $\phi(y)=
\varphi(g(y))$.  The action of the induced representation on $\phi$
can be deduced from the above transformation rule to be
$$U(g_1)\phi(y) = D(h(y,g_1))\phi (y^\prime)  ,\  \forall\  g_1\ \in\
G
\eqno(3)$$
where
$$yg_1= h(y,g_1) y^\prime.
\eqno(4)$$
The infinitesimal form of this
equation is given by
$$\delta \phi = V \phi + D(\delta h)\phi,
\eqno (5)$$
where $V= \delta y^M {\partial \over \partial y^M}$, $\delta y^M$ are
the transformations of the coordinates and the second term  is
determined from equation  (4) by taking $g_1=1+\delta g_1$
and   expanding up to first order in variations.
Hence, when defined on
${G\over H}$ the fields $\phi$ have a tensor character corresponding to the
above isotropy group transformations. The mapping between coordinate
patches on ${G\over H}$ can be realised by a coordinate transformation
which is induced by the right action of a group element on ${G\over H}$ and
consequently, under this transformation $\phi$ under goes a $H$
matrix transformation in accord with equation (3).
\par
For $N=2$ harmonic superspace we denote the group element of
$ SU(2)$ by $u^{\ i}_I,\ i=1,2$, $I= +,-$. The hypermultiplet
can be described by a superfield on $\varphi_+$ on $M_2\times SU(2)$
 which satisfies the   constraints
$$ D_{0}\varphi_+=\varphi_+,\ D_{ +}^{\ -}\varphi_+=0,
\eqno( 6)$$
$$D_{\alpha +}\varphi_+=0,\ \ D_{\dot \alpha }^{\ -}\varphi_+=0,
\eqno (7)$$
In the first equation  $D_+^{\ -},D_-^{\ +}$ and $D_0$ are the
right-invariant vector fields   on $SU(2)$ which are the same as the
vector fields induced by the left action of $SU(2)$  acting on itself
and
$D_0$ generates the action of the isotropy group, $U(1)$.
In the second equation, $D_{\alpha I}= u_I^{\ j}
D_{\alpha j},\  D_{\dot \alpha }^I= u_j^{\ I} D_{\dot \alpha } ^{\ j},\
j=1,2;\ I= \{+,-\}$ where $u_i^{\ I}$ is the inverse of $u_I^{\ i}$.
\par
The first condition of the equation (6) is equivalent to the
equivariant condition discussed above provided  $\varphi_+$ has $U(1)$
charge 1, while the second condition implies that $\varphi_+$ is
 a holomorphic "function" on $S^2= {SU(2)\over U(1)}$.
The constraints of equation (7) are
referred to as Grassmann analyticity constraints [13] and by expanding
$\varphi_+$ in harmonic variables we find  that they  imply the relation
$\varphi_+= u_+^{\ i}\varphi_i$ where $\varphi_i$ is the usual $N=2$ matter
field defined on  $M_2$ which obeys the constraints of equation (1).
The  conjugate superfield
$\tilde \varphi_+$ satisfies the  identical  constraints of equations
(6)  and (7) which imply that it has the form
$\tilde \varphi_+= u_+^{\ i}\bar \varphi_i$.
\par
In the  $N=4$ case, we write elements $u\ \in SU(4)$ as
$ u_I^{\ i}=( u_r^{\ i}, u_{r\prime}^{\ i})$, $r=1,2,\ r^\prime =3,4,\
i=1,\ldots ,4$. The $N=4$ Yang-Mills theory field strength is
a superfield $W$ defined on $M_4\times SU(4)$
 subject to the Grassmann analyticity constraints
$$ D_{\alpha r} W=0,\  D_{  \dot \alpha}^{\ r ^\prime} W=0,\
\eqno (8)$$
 as well as the constraints
$$ D_{r}{}^{s^\prime } W=0,\  D_{0} W=2W.
\eqno(9)$$
 The super field $W$ also satisfies  a reality condition which is
equivalent to the self-duality of $W_{ij}$. The above spinor covariant
derivatives are defined analogously to the
$N=2$ case, and the right-invariant vector fields on $SU(4)$ are
$D_{r}^{\ s}$,
$D_{r^\prime }^{\ s^\prime}$, $D_{0}$ and $D_{r}^{\ s^\prime}$,
$D_{r^\prime }^{s}$ where the former correspond to the isotropy group
and the latter the coset.  The superfield $W$ is a singlet under
$SU(2)\times SU(2)$, but has $U(1)$ charge 2. The first condition of
equation (9) states that $W$ is  a holomorphic "function" on
$ {SU(4)\over S(U(2)\times U(2))}$ and the $U(1)$ constraint implies [14] that
$W= \epsilon^{rs} u_r^{\ i} u_s^{\ j} W_{ij}$ where $W_{ij}$ is a
superfield on Minkowski superspace that satisfies the
constraints of equation (2).
\par
In the above, we have  worked with
superfields defined on the (internal) group, however,  for the applications in
this
paper it will be easier to work with superfields defined directly on
the coset spaces as discussed above. We now briefly describe how this
works for the hypermultiplet.  The group manifold $SU(2)$ can be
covered by  two patches\footnote*{The patches correspond to the patches of
$S^2$; strictly one should introduce two patches for the $U(1)$ fibres as
well.} $U$ and $U^\prime$ and in $U$ we
parameterise  group elements by
$$ u_I^{\ j} = { 1 \over \sqrt {(1+y \bar y)}}\left(\matrix{
e^{i\vartheta}&0            \cr
0          & e^{-i\vartheta}\cr
}\right)
\left(\matrix{
1        &y            \cr
-\bar y   & 1          \cr}\right)   .
\eqno (10)$$
 and in $U^\prime$  by an similar form with
$y^\prime$ and $\vartheta^\prime$ replacing $y$ and $\vartheta$. On $U$ the
right-invariant
vector fields on $SU(2)$ are given by
$$D_-^{\ +}=(1+y \bar y)e^{-2i\vartheta}\big( {\partial \over
\partial y} +{i\over 2}{\bar y \over (1+y \bar y)} {\partial \over
\partial \vartheta}\big ),
\ D_0=-i{\partial \over \partial \vartheta},
$$
$$ D_+^{\ -} = (1+y \bar y)e^{2i\vartheta}\big( {\partial \over
\partial \bar y} -{i\over 2}{ y \over (1+y \bar y)} {\partial \over
\partial \vartheta}\big ).
\eqno(11)$$
On the coset ${SU(2)\over U(1)}$, we choose  our coset representatives to be
of the above form with $\vartheta =0$ and $\vartheta^\prime  =0$ in the
corresponding two patches, also denoted $U$ and $U^\prime$, on
$SU(2)/U(1)$. The transformation between the two patches is given by
$y \to y^\prime ={1\over y}$  and it is a straight forward exercise
to compute, in term of the parameterisation of equation (10), the group
transformation which takes the coset representatives in each patch
into each other. One finds that it involves an isotropy rotation for
which
$e^{2i\vartheta } ={ y \over \bar y}$ and hence the induced
representation is related on the two patches by
$$ \phi^\prime (y^\prime , \bar y^\prime ) =  {({ y \over
\bar y})}^{q\over2}\phi(y,\bar y)
\eqno(12)$$
if $\phi $ has $U(1)$ charge q, i.e. $\phi \to e^{iq\vartheta}\phi  $.
\par
In this coset formulation, the constraints of equations (6) and (7)
for $\varphi_+$ have a simple interpretation. The $D_0$ constraint
tells us that
$\phi_+$ has
$U(1)$ charge 1. On making the change
$\phi_+ \to { 1\over \sqrt {(1+y\bar y)}} \phi_+$
we finds that the last condition of equation (6) states that
$\phi_+$ is independent of $\bar y$.  Consequently,
$\phi_+$ depends  on  $y$ alone in the patch $U$ and $y^\prime$ in
the patch $U^\prime$. The  field $\phi_+$ is assumed to be
well defined on all of
$SU(2)\over U(1)$. It is well defined in the
patches $U$ and $U^\prime$ if $\phi_+$ is polynomial in
$y$ around $y=0$ and polynomial in ${ y^\prime  }$ as $y^\prime
\to 0$ respectively. In order for $\phi_+ $ be globally defined, we
must ensure the matching of these two conditions as we change
from $U$ to $U^\prime$. Using equation (12)  we find that $\phi_+$
 can only be of the form
$\phi_+= \phi_1 + y \phi_2$ where $\phi_1$ and $\phi_2$ are
independent of $y$  and
 are the components of the hypermultiplet in it's original Minkowski
space formulation  formulation. Thus we recover the above relation
between the two formulations,   namely $\phi_+ = u_+{}^i \phi_i$.
\par
The spinor  constraints of equation (7) imply that $\phi_+$ and
$\tilde \phi_+$ depend only on  the variables
$$\lambda ^{\alpha } \equiv \theta ^{\alpha 2}-y \theta ^{\alpha 1},
\  \pi ^{\dot \alpha } \equiv \theta ^{\dot \alpha} _1 + y  \theta
^{\dot \alpha }_{2},$$
$$  s^{\alpha \dot \alpha}\equiv
x^{\alpha \dot \alpha} +{i\over 2}\theta _1^{ \alpha }(\theta
_1^{\dot \alpha }+y \theta _2^{\dot \alpha })
- {i\over 2}(\theta ^{\alpha }_2 - y \theta ^{\alpha }_1)
\theta ^{\dot \alpha }_2 \ \ {\rm and}\ \ y\ .
\eqno(13)$$
 Hence, by working on the coset, we can solve all the
constraints of equations (6) and (7) and work with fields that only
depend on the above coordinates and are polynomials in $y$ of degree at
most one.  Such superfields can be regarded as functions
on a new superspace, called analytic superspace [13] which we now
discuss.
\par
Harmonic superspaces, like ordinary superspace can be regarded as
coset spaces of the super Poincar\'e groups. They can
also be viewed as coset spaces of superconformal groups [17]. In this
paper, we will be concerned with  superconformal theories and so we
will adopt this latter point of view. In particular, the theories of
interest to us involve chiral and analytic
superfields  and these are best described using chiral and analytic
superspaces. However, these superspaces are coset spaces of the
complexified superconformal group which is $SL(4|N;C)$ where $N$ is
the number of supersymmetries. The relevant isotropy groups are of
parabolic type, meaning that, if we regard an element of $SL(4|N;C)$ as
made up of four blocks in the obvious way,   each block should
itself be of block lower triangular form in order for the element to
belong to the isotropy group [17]. For example  $N=2$ analytic
superspace has an isotropy group consisting of elements of
$SL(4|2;C)$ of the form
$$ \left(\matrix{
\times&\times&      &      &\times&      \cr
\times&\times&      &      &\times&      \cr
\times&\times&\times&\times&\times&\times\cr
\times&\times&\times&\times&\times&\times\cr
\times&\times&      &      &\times&      \cr
\times&\times&\times&\times&\times&\times\cr
}\right),
\eqno (14)$$
where $\times$ represents a non-zero
entry. The coset representative for this superspace can be chosen
to be of the form
$$ \left(\matrix {
I&-is      & 0& -i\lambda \cr
0&I        & 0&0          \cr
0& -i\pi   &1 & y         \cr
0&0        &0 &1          \cr
}\right)
\eqno (15)$$
where we have the index structure $s^{\alpha\dot{\alpha}}$, $\lambda^{\alpha}$
and $\pi^{\dot
\alpha }$.
\par
The $N=4$ analytic
superspace which relevant for $N=4$ Yang-Mills has an isotropy group
consisting of elements of
$SL(4|4;C)$ of the form
$$ \left(\matrix{
\times&\times&      &      &\times&\times&       &      \cr
\times&\times&      &      &\times&\times&       &      \cr
\times&\times&\times&\times&\times&\times&\times &\times\cr
\times&\times&\times&\times&\times&\times&\times &\times\cr
\times&\times&      &      &\times&\times&       &      \cr
\times&\times&      &      &\times&\times&       &      \cr
\times&\times&\times&\times&\times&\times&\times &\times\cr
\times&\times&\times&\times&\times&\times&\times &\times\cr
}\right),
\eqno (16)$$
In fact, the superconformal group of $N=4$ is not $SL(4|4;C)$, but
rather the simple supergroup obtained from it by factoring out the subgroup
consisting of elements proportional to the identity matrix. The coset
representative for this  analytic  superspace can
be chosen to be of the form
$$ \left(\matrix {
I&-is      & 0& -i\lambda \cr
0&I        & 0&0          \cr
0& -i\pi   &I & y         \cr
0&0        &0 &I          \cr
}\right)
\eqno (17)$$
where  $y$ is a two by
two matrix whose index structure will be denoted by $y^{a a^\prime}$.
Similarly, we have $\lambda ^{\alpha a^\prime }$ and $\pi^{a \dot
\alpha }$. We note that the bottom right hand corner of the isotropy
group elements is the same as the top left corner.
The above spaces have compact bodies whereas we wish eventually to
consider non-compact Minkowski space. However, there is a
correspondence between analytic superspace and Minkowski superspace
via harmonic superspace which allows one to restrict  to the usual
non-compact super Minkowski space, although the internal spaces
remain compact [17]. This step is analogous to
removing the point at infinity on the Riemann sphere.
\par
We now consider induced representations on the above coset spaces.
The fields which carry these representations depend only on
$s,\lambda,\pi$ and $y$, and  are holomorphic.
The $N=2$ matter and $N=4$ Yang-Mills multiplets correspond to
particular representations of this type and in this framework are not
subject to any constraints beyond the requirement of holomorphicity,
meaning they  are non-singular. This makes contact with the discussion
of these multiplets given above where they were defined on the real
coset spaces $\hat M_N,\ N=2,4$  subject to the constraints of
equations (6) and (7) and equations (8) and (9) respectively. As we
have illustrated for the hypermutiplet we can solve the constraints
and so recover the analytic coset space with the coordinates
 of equation (13). For the purposes of this
paper it will be simpler to adopt the analytic superspace viewpoint from
the outset.
\par
Given a superconformally  invariant supersymmetric
theory, its Green's
functions will obey   Ward identities corresponding to the
superconformal transformations.  Given such a n point Green's
function
$G$ for  superfields $\phi_ p, p=1,\ldots,n$ of weight
$q_p$, then the Ward identities are of the form
$$ \sum _p\big \{ V_M^p + {q_p\over N } \Delta ^p_{M} \big\}G=0.
\eqno (18)$$
where, using equation (5) we identify
${q\over N}\Delta_M \phi \equiv D(\delta h_{M 1})\phi$. The subscript
$M$ labels the particular conformal transformation. For chiral and
analytic superfields these Ward identities take a particularly simple
form when expressed in terms of the coordinates of the corresponding
chiral and analytic superspaces.
\par
We begin by considering the chiral case first and  we shall argue that
at  a fixed point of  any supersymmetric theory
  the Green's function of gauge invariant chiral operators of the
same chirality are determined by superconformal invariance.
As explained above,  a chiral superfield $\phi$ is only   a function
of  $s^{\alpha\dot \alpha}$  and $\theta^{\alpha i}$.
A Green's function G of n chiral
superfields will be a function on n
copies of chiral superspace, a space which has coordinates $s^{\alpha\dot
\alpha}_p,\
\theta^{\alpha i}_p,\ p=1,\ldots ,n$. The superconformal
Ward-identities for translations, dilations and  special conformal
transformation are
$$\sum  \{\partial_{\alpha \dot \alpha} \big \} G=0,
\eqno (19.1)$$
$$\sum  \big \{s ^{\alpha \dot \alpha }
\partial_{\alpha \dot
\alpha}
 + {1\over 2} \theta ^{\alpha j} \partial_{\alpha j }
+ q{(4-N)\over N} \big \} G=0,
\eqno (19.2)$$
$$\sum  \big \{s ^{\alpha \dot \beta }C_{\dot \beta \beta}
s ^{\beta \dot \alpha } \partial_{\alpha \dot \alpha}
+s ^{\alpha \dot \beta }C_{\dot \beta \beta} \theta ^{\beta j}
\partial_{\alpha j }
+ q{(4-N)\over N}s^{\beta \dot \beta} C_{\dot
\beta \beta} \big \}G=0,
\eqno (19.3)$$
those for supersymmetry are
$$\sum \{ {\partial_{\alpha i} }\} G=0,
\eqno(19.4)$$
$$\sum \{ \theta^{\alpha i} {\partial_{\alpha \dot \alpha} } \}G =0 .
\eqno (19.5)$$
For the internal symmetry, with traceless parameter $ E_j^{\ i}$, we
have the corresponding Ward identity
$$ \sum\{ \theta ^{\alpha j} E_j^{\ i} \partial _{\alpha i} \}G=0.
\eqno (19.6)$$
and finally those for S-supersymmetry are given by
$$ \sum  \big \{s ^{\beta \dot
\alpha  }\partial _{ \dot\beta i}  \}G=0,
\eqno (19.7)$$
$$\sum\{ s ^{\beta \dot\alpha } \theta ^{\alpha i}\partial _{\alpha
\dot \alpha}
+\theta ^{\beta i}  \theta ^{\alpha j}\partial_{\alpha j}
+q{(4-N)\over N}\theta ^{\beta j} \big \}G=0.
 \eqno (19.8)$$
In the above equations the sum is over $p$, however, this index is
suppressed on the coordinates and we have used the shorthand notation
$\partial_{\alpha \dot \alpha}={\partial \over \partial s^{\alpha
\dot \alpha}}$ and
$\partial _{ \alpha i}= {\partial \over \partial \theta
^{\alpha  i}}$.
For $N\neq 4$ we also have R symmetry with the corresponding
Ward identity
$$ \sum \{\theta ^{\alpha j} {\partial_{\alpha j}} -2q \}G=0.
\eqno (20)$$
\par
Translation invariance, encoded in  equation (19.1), implies that
$G$ is a function of
$s^{\alpha \dot \alpha}_{pq}= s^{\alpha \dot \alpha}_p - s^{\alpha
\dot \alpha}_q$, while  the first supersymmetry constraint  on $ G$
implies that it is a function of $\theta^{\alpha
i}_{pq}=\theta^{\alpha i}_p - \theta^{\alpha i}_q$. Let us now consider  a
superconformal chiral invariant f, which will satisfy all the
superconformal
Ward identities with no isotropy group transformations. It will be a
function of $s^{\alpha \dot
\alpha}_{pq}$ and   $ \theta^{\alpha i}_{pq}$. The R symmetry implies
that $f$ will satisfy  the constraint
$$\sum \{ \theta ^{\alpha j} {\partial \over \partial \theta^{\alpha
j}}\}f=0.
\eqno (21)$$
which implies that it is independent of $\theta ^{\alpha i}$.
The last supersymmetry constraint of equation (19.5)
implies that $f$ is independent of $s^{\alpha \dot \alpha}_{pq}$ and so
is a constant. Hence, we have shown that there are no
superconformal chiral invariants except for a constant.
Although R symmetry implies that any solution to
the superconformal Ward identities for a  given
chiral Greens functions has the same power of
$\theta$, it is not a legitimate step to divide such
nilpotent quantities and so one can not conclude that
chiral Greens functions can not contain arbitary functions.
\par
We now consider analytic superfields of $N=2$ matter and $N=4$
Yang-Mills. The transformations of these fields are found
by using the coset construction discussed above. After some
calculation, one finds, for the case $N=4$, that the results for
translations T , dilations D and special conformal transformations K
are
$$\delta_T \phi =  \big \{\partial_{\alpha \dot \alpha} \big \} \phi,
\eqno (22.1)$$
$$\delta_D \phi =  \big \{s ^{\alpha \dot \alpha }\partial_{\alpha \dot
\alpha}
 + {1\over 2} \lambda ^{\alpha a^\prime} \partial_{\alpha a^\prime }
+ {1\over 2} \pi^{a \dot \alpha }\partial _{a \dot \alpha}
+ {q\over 2} \big \} \phi, \eqno (22.2)$$
$$\delta_K \phi =  \big \{s ^{\alpha \dot \beta }C_{\dot \beta \beta}
s ^{\beta
\dot \alpha }
 \partial_{\alpha \dot \alpha}
+s ^{\alpha \dot \beta }C_{\dot \beta \beta} \lambda ^{\beta a^\prime}
\partial_{\alpha a^\prime }
+  \pi^{a \dot \beta } C_{\dot \beta \beta} s ^{\beta \dot \alpha}
\partial _{a \dot \alpha}
$$
$$-i \pi^{a \dot \beta } C_{\dot \beta \beta} \lambda ^{\beta
a^\prime}
\partial _{a a^\prime}
+ {q\over 2}s^{\beta \dot \beta} C_{\dot \beta \beta} \big \} \phi,
\eqno (22.3)
$$
 those of supersymmetry are
$$ \delta_{S_{\alpha a^\prime}} \phi =  \big \{
\partial_{\alpha a^\prime } \big \}\phi, \eqno (22.4)$$
$$ \delta_{S_{a \dot \alpha }} \phi =  \big \{ \partial _{a \dot
\alpha}
 \big \}\phi, \eqno (22.5)$$
$$ \delta_{S_{\alpha}^b} \phi =  \big \{- y^{b a ^\prime}
\partial_{\alpha a^\prime } +i  \pi^{b \dot \alpha }\partial_{\alpha
\dot \alpha}\big \} \phi, \eqno (22.6)$$
$$ \delta_{S_{ \dot \alpha }^{b^\prime}} \phi =   \big \{y^{a b
^\prime}
\partial _{a
\dot \alpha} +i\lambda ^{\alpha b^\prime}\partial_{\alpha \dot \alpha}
\big \}\phi, \eqno (22.7)$$
those of internal transformations are
$$ \delta_{I_{a a^\prime}} \phi =  \big \{\partial_{a
a^\prime}
\big \}\phi, \eqno (22.8)$$
$$ \delta_I \phi =   \big \{y^{a a ^\prime} \partial_{a a^\prime}
+{1\over 2}\lambda ^{\alpha a^\prime}\partial_{\alpha a^\prime }
 +{1\over 2}\pi^{a \dot \alpha } \partial _{a \dot \alpha}
-{q\over 2}
\big \}\phi, \eqno (22.9)$$
$$ \delta_{I^{b^\prime b}} \phi =   \big \{-y^{a b ^\prime} C_{
b^\prime b} y^{b a ^\prime}
\partial_{a a^\prime}
 -\lambda ^{\alpha b^\prime}C_{ b^\prime b} y^{b a ^\prime}
\partial_{\alpha a^\prime }
- y^{a b ^\prime} C_{ b^\prime b} \pi^{b \dot \alpha }\partial _{a \dot
\alpha}
$$
$$+i\lambda ^{\alpha a^\prime} C_{a^\prime b}  \pi^{b \dot \alpha }
\partial_{\alpha\dot \alpha}
+{q\over 2} y^{b b ^\prime} C_{ b^\prime b}
\big \}\phi,  \eqno (22.10)$$
while those of special supersymmetry are
$$ \delta_{SS^\beta_a} \phi =  \big \{is ^{\beta \dot
\alpha  }\partial _{a
\dot
\alpha}  +\lambda ^{\beta a^\prime}\partial_{a a^\prime} \big \}\phi,
\eqno (22.11)$$
$$ \delta_{SS^{\dot \beta}_{a^\prime}} \phi = \big \{ -is ^{\alpha
\dot \beta }\partial_{\alpha a^\prime } + \pi^{a \dot \beta
}\partial_{a a^\prime}
\big \}\phi, \eqno (22.12)$$
$$ \delta_{SS^{\dot \beta b}} \phi =  \big \{ -i s ^{\alpha \dot \beta
} y^{b a^\prime}\partial_{\alpha a^\prime } + \pi^{a \dot \beta } y^{b
a^\prime}\partial_{a a^\prime}  - s ^{\alpha \dot \beta }\pi^{b \dot
\alpha}\partial_{\alpha \dot
\alpha}
 + \pi^{a \dot \beta } \pi^{b \dot \alpha }\partial _{a \dot
\alpha}
+ {q\over 2}\pi^{b \dot \beta } \big \}\phi, \eqno (22.13)$$
$$\delta_{SS^{ \beta b^\prime}} \phi =  \big \{ iy^{a b ^\prime}
s ^{\beta \dot
\alpha }\partial _{a \dot \alpha}  + y^{a b ^\prime} \lambda ^{\beta
a^\prime}\partial_{a a^\prime}  -\lambda ^{\alpha b^\prime} s ^{\beta
\dot \alpha }\partial_{\alpha
\dot \alpha}
-\lambda ^{\alpha b^\prime} \lambda ^{\beta
a^\prime}\partial_{\alpha a^\prime }
-{q\over 2}\lambda ^{\beta b^\prime}
\big \}\phi, \eqno (22.14)$$
In the above equations we have used the shorthand notation
$\partial_{\alpha \dot \alpha}={\partial \over \partial s^{\alpha
\dot \alpha}}
$,
$\partial _{ \alpha a^\prime}= {\partial \over \partial \lambda
^{\alpha a ^\prime}}$,
$\partial _{a \dot \alpha}= {\partial \over \partial \pi ^{a \dot
\alpha}}$ and
$\partial_{a a^\prime}=  {\partial \over \partial y^{a a^\prime}}$.
These transformations are for $N=4$ analytic  superfields which have
$U(1)$ charge $q$ where the $U(1 )$ is the obvious $U(1)$ in the
$S(U(2)\times U(2))$ isotropy group. When using the
constrained superfields this is equivalent to the constraint
 $D_0\phi=q\phi$.
\par
We can obtain the superconformal transformations for an $N=2$
analytic superfield of charge   $U(1)$  q (i.e. $D_0\phi=q\phi$)
by making the following substitutions in equation (22)
$$\lambda ^{\alpha a ^\prime } \to
\lambda ^{\alpha}, \ \ \pi ^{ a \dot \alpha} \to \pi^{\dot \alpha},
\ \ y^{a a^\prime} \to y\
 \eqno(23)$$
and ${q\over 2}\to q$ in
equations (22.1), (22.2) and (22.3).
We must also add the R transformations which are of the
form
$$\delta_R \phi =  \big \{ \lambda ^{\alpha } \partial_{\alpha } -
 \pi^{a \dot \alpha }\partial _{a \dot \alpha}
\big \}\phi
\eqno(24)$$
since such transformations are not present for $N=4$.
\par
Observables in rigidly supersymmetric gauge theories divide into
two types, non-local  observables such as  Wilson loops and  local
gauge invariant operators constructed from products of the elementary
fields of the theory. Our aim will be to calculate Green's functions of
these latter observables using the superconformal Ward identities of
equation (18) and the transformations of equation (22).  For
$N=1$ supersymmetric theories these local observables can only be
constructed from $W_\alpha, \phi$, and their conjugates which are
anti-chiral. As we have seen above, the superconformal Ward identities
only determine Green's functions in the  chiral or anti-chiral sector.
For $N=2$ theories,  the basic superfields  are the Yang-Mills
superfield strength $W$ and its conjugate
$\bar W$  and the N=2 matter superfield
$\phi_+$ and its conjugate $\tilde \phi_+$. A very important point for
our analysis is that while the $N=2$ Yang-Mills
 superfield $W$ is chiral, its complex conjugate $\bar W$ has
constraints of the opposite chirality, while the superfield for N=2
matter and its complex conjugate have the same chiral constraints. As
we shall see the superconformal Ward identities  only
determine  Green's function of operators in a given chiral or
analyticity sector. However, since N=2 matter has the same analyticity
properties as its conjugate we  can hope to restrict all such
Green's function involving any gauge invariant $N=2$ matter, although
we can still only determine the Green's functions of the chiral or
anti-chiral sectors of gauge invariant products of  the
$N=2$ Yang-Mills superfield.
\par
Although the $N=4$ Yang-Mills multiplet  is described by a real
single analytic superfield $W$, it is not the case that all composite
multiplets can be obtained as products of $W$. To see this consider the
example of $TrW_{ij}W_{kl}$; this product decomposes into a
\underbar {20} and a \underbar {1}
under
$SU(4)$, but only the \underbar {20} is present in $TrW^2$ as may easily be
checked. In the harmonic formalism, we can define a superfield
$W^\prime =\epsilon^{r^\prime s^\prime} u_{r^\prime}{}^i
u_{s^\prime}{}^j W_{ij}$  which is anti-analytic. The product
$TrWW^\prime$ includes both the \underbar {20} and the \underbar {1}.
We can hope to  use the
superconformal Ward identities to determine all Green's functions of
gauge invariant polynomials of the superfield $W$.
\par
To get a feel for the way the superconformal  Ward
identities act, we now calculate the two point function, $G_{12}$ in
$N=4$ Maxwell theory. This is a function of
$s_p^{\alpha\dot \alpha}$,
$\lambda_p^{\alpha a^\prime }$,
$\pi^{a \dot \alpha}_p$ and $y^{a a^\prime}_p,\ p=1,2$.
Taking  the transformation equation to mean also the corresponding
Ward identity, equations (22.1), (22.4),(22.5),(22.8) respectively
imply that $G_{12}$ is a function of their differences
$s^{\alpha\dot
\alpha }= s_1^{\alpha\dot \alpha}-s_2^{\alpha\dot \alpha}$,
$\lambda^{\alpha a^\prime }= \lambda_1^{\alpha a^\prime }
- \lambda_2^{\alpha a^\prime }$,
$\pi^{a \dot \alpha} =\pi^{a \dot \alpha}_1 -\pi^{a \dot
\alpha}_2$ and $y^{a a^\prime}= y^{a a^\prime}_1- y^{a
a^\prime}_2$. The Green's function consisting of two $W$'s
,which have $q=2$ corresponding to the fact that they have dilation
weight one, has its form fixed by equation (22.2) and Lorentz
invariance to be given by
$$G_{12} \equiv <W(1) W(2)> = {h(y)\over s^2} + {f(y)\over {(s^2)}^2}
\lambda^{\alpha a^\prime}\pi^{a \dot \alpha}
s_{\alpha \dot \alpha}y_{a a^\prime}+\ldots,
\eqno(25)$$
where $+\ldots$
denote terms of higher order in anti-commuting variables, $s^2=
s^{\alpha \dot \alpha}s_{\alpha \dot \alpha}$ and h and f are
functions of $y^{a a^\prime}$. Equation (22.9) implies that
$f\propto y^0$
and $h\propto y^2$ which taken with the other internal identities
imply that $h= y^2\equiv y^{a a^\prime}y_{a a^\prime} $ and $f$ is a
constant. The Green's function can now be obtained, for example,  by
using equation (22.6) and the final result can be written in the form
$$G_{12}= {\hat y^2\over s^2},\eqno(26)$$
where $\hat y^{a a^\prime }= y^{a a^\prime }
-2i{ \lambda^{\alpha a^\prime}\pi^{a \dot \alpha}  s_{\alpha \dot
\alpha}\over s^2}$.
Since one does not have $R$ invariance in $N=4$ Yang-Mills theory one
can in  principle have a $\lambda^2$ terms, however, one can verify
that no such terms are possible for the two point function.
\par
One can carry out an analogous calculation for the two point function
for $N=2$ matter, the result is of the form of equation (26) with the
replacements of equation (23).
\par
We can construct two point and higher point
 Green's functions in $N=4$ Yang-Mills by taking product of appropriate
powers of the above  two point function. These will also satisfy the
superconformal identities of equation (18), since these contain
operators that obey the Leibnitz rule. As we are interested in gauge
invariant operators we consider $TrW^n$ which obeys the same
constraints as $W$ of equations (8) and (9) except that $D_0TrW^n=2nTr
W^n$. As a result, we may satisfy the superconformal Ward identities by
taking
$$<TrW^{n_1}(1)\ldots TrW^{n_N}(N)>= \prod_{i<j}
{G_{ij}}^{{1\over (N-2)}(n_i +n_j -{n\over N-1})},
\eqno(27)$$
where $n=\sum n_i$.
\par
The above  expression is not unique as we can form the
invariants such as
$$\hat u = {G_{13}G_{24}\over G_{12}G_{34}},\
\hat v = {G_{14}G_{23}\over G_{12}G_{34}},
\eqno(28)$$
and  we can multiply the above expression by an arbitrary function of
these and still obtain Green's functions that satisfy the
superconformal  Ward identities. However, one must take care to
ensure that analyticity in $y^{a a^\prime}$ is preserved. Since the
highest  power  of $y^{a a^\prime}$ is encoded by the internal
identities of equation (22.9), this condition means that one should
have Green's functions which have no poles in
$y^2$. As a result, all possible Green's functions constructed in
this way are products of two point functions.  The above invariants
can be written in the form
$\hat u = {u\over
\hat t},\
\hat v = {v\over \hat w}$ where $u= {s^2_{12}s^2_{34}\over
s^2_{13}s^2_{24}}$,
$v= {s^2_{12}s^2_{34}\over s^2_{14}s^2_{23}}$, $\hat t =
{\hat y^2_{12} \hat y^2_{34}\over \hat y^2_{3}\hat y^2_{24}}$,
$\hat w= {\hat y^2_{12}\hat y^2_{34}\over \hat y^2_{14}\hat
 y^2_{23}}$.
\par
 Any Green's function in $N=4$ Yang-Mills theory can be written as the product
of two point functions that appear in equation (27) times a superconformal
invariant. In particular, the four point Green's fnction can be written as the
product of $G_{12}$ and  $G_{34}$ to suitable powers
times a superconformal invariant depending on four points in analytic
superspace. Hence, the first step in determining the four point Green's
function is
to find what the four point superconformal invariants are.  Clearly, any
function of the
$N=4$ analytic  invariants, $\hat u,\ \hat v$ is an invariant, but
we can ask if these the only independent invariants.
  Invariants  are quantities which obey all the
superconformal Ward identities of equation (18) with $q=0$, but they need not
be analytic functions of the $y_{ij}$.  An invariant can only be of the form
$$ I= f + \lambda^{\alpha a^\prime}_{mn}\pi^{a \dot \alpha}_{pq}
H_{\alpha \dot \alpha ; a a^\prime }^{mn;pq} +\ldots
\eqno(29)$$
where we have already used the fact that the invariant must be  a
function of differences and  $mn=12,23,34$. Equations (22.3) and
(22.10) imply that f can be regarded  as only a  function of $\hat u,\
 \hat v, t$ and
$ w$.  The strategy is to find the $y$ and $s$ dependence of those
$H^{mn;pq}$'s which involve $mn$ and  $pq=12$ or $34$ using the
non-linear internal and non-linear special conformal Ward identities
respectively. We then solve for the remaining  components of
$H^{mn;pq}$ using the Q-supersymmetry identities and then finally use
 S-supersymmetry to show that $H=0$ and that f does not depend on
$t $ and $w$. This calculation is very lengthy and we only give the
briefest outline. To illustrate the calculation we carry it
out making the assumption that $H_{\alpha \dot \alpha ; a a^\prime }^{mn;pq} $
can be expressed in terms of $(s_{ij})_{\alpha \dot \alpha }$times coeficients
that depend on the values of  $a, a^\prime,m,n,p$ and $ q$. Using equations
(22.3) and (22.10) we find that
$$ H^{12 :12 }_{\alpha \dot \alpha ; a a^\prime}
 = \sum _{pq}{{(y_{1p})}_{a a^\prime }\over y^2_{1p}}
{{(s_{1q})}_{\alpha \dot \alpha  } \over s^2_{1q}} h^{12,12}_{1q;1p},
\  H^{ 12:34 }_{\alpha \dot \alpha ; a a^\prime}
 = {{(y_{14})}_{a a^\prime }\over y^2_{14}} {{(s_{14})}_{\alpha
\dot \alpha  }\over s^2_{14}} h ^{12,34},$$
$$  H^{34 : 34}_{\alpha \dot \alpha ; a a^\prime}
 = \sum _{pq} {{(y_{p4})}_{a a^\prime } \over y^2_{p4}}
{{(s_{q4})}_{\alpha
\dot \alpha  }\over s^2_{q4}} h ^{34,34}_{q4;p4},
\  H^{ 34:12 }_{\alpha \dot \alpha ; a a^\prime}
= {{(y_{14})}_{a a^\prime } \over y^2_{14}} {{(s_{14})}_{\alpha
\dot \alpha  }\over s^2_{14}} h ^{34,12}.
\eqno(30)$$
where the $h$'s depend on only $\hat u,\ \hat v,\ t$ and $w$.
We now use equations (22.6) and (22.7) to solve for the remaining
components of $H$. For example, one finds that
$$ H^{12;23}_{\alpha \dot \alpha ; a a^\prime} =
\sum _p{{(y_{13})}_{a a^\prime }\over y^2_{13}}
{{(s_{1p})}_{\alpha \dot \alpha  }\over s^2_{1p}} h ^{12,12}_{1p;13}
+ {{(y_{14})}_{a a^\prime } \over y^2_{14}} {{(s_{14})}_{\alpha
\dot \alpha  } \over s^2_{14}} h ^{12,34}.
\eqno(31)$$
Finally, substituting equations (30) and (31) into (22.12) one
discovers that $H^{12;12}_{\alpha \dot \alpha ; a a^\prime}= 0=
H^{34;12}_{\alpha \dot \alpha ; a a^\prime} = {\partial f\over \partial
t}=  {\partial f\over \partial w} $. Solving for the other components
of  $H$ from the supersymmetry identities and using equation (22.11)
one finds that all the components of $H$ vanish.
\par
 This result illustrates the strength of the superconformal
Ward identities. However, the above assumption on the form of $H_{\alpha \dot
\alpha ; a a^\prime }^{mn;pq} $ is not the most general since for fixed
values of $a, a^\prime,m,n,p$ and $ q$ , $H_{\alpha \dot \alpha ; a a^\prime
}^{mn;pq} $ can take on four values
while $(s_{ij})_{\alpha \dot \alpha }$ has only three values $(s_{12})_{\alpha
\dot \alpha }$, $(s_{23})_{\alpha \dot \alpha }$ aqnd $(s_{34})_{\alpha \dot
\alpha }$. The most general form of $H_{\alpha \dot \alpha ; a a^\prime
}^{mn;pq} $ would include a term of the form
$(s_{34}s_{23}s_{12})_{\alpha \dot \alpha }$. This is equivalent to including
the term $\epsilon ^{\mu\nu\rho\tau} (s_{12})_\nu (s_{23})_\rho \times$ 
$(s_{34})_\tau
(\sigma_\mu)_{\alpha \dot \alpha }$. The full calculation including this term
is too complicated to present in this paper, but one finds that in addition to
$\hat u,\ \hat v$ there are two other invariants.
Their precise form will be given in reference [18].
\par
Although preliminary calculations suggest otherwise, one
could, in principle,  have four point invariants which begin with
${(\lambda \pi)}^2$ or higher powers. However, invariants  of this type
can only  involve functions  of $ \hat u$ and $\hat v$ as a consequence of the
above argument.  Since R symmetry is not a symmetry of $N=4$ Yang-Mills
theory it might seem that there could be terms of the form
$\lambda^2 +\cdots$  etc. R transformations act only on the spinor coordinates
by $ \lambda \to
a \lambda ,\   \pi \to \bar{a} \pi $, where $|a|=1$: it commutes with the
superconformal transformations and as a result, the
superconformally invariant Green's functions can be classified
according to their R weights and calculated separately. However,  $N=4$
Yang-Mills theory is invariant under the $Z_4$ subgroup of R
transformations (i.e. $a^4=1$) under which the $N=4$
Yang-Mills field strength transforms as $W\to - W$.  This symmetry
forbids, for example,   the occurrence  of  $\lambda^2 +\cdots$ terms in   the
four point invariant.
\par
 The function of the invariants appearing in a Green's function is not
arbitrary, but must be such that the
Green's function is an analytic function of $(y_{ij})_{a a^\prime}$.
This places very strong constraints on the form of the function of the
invariants and we believe that for four point
Green's functions composed of a suitable class of
operators analyticity is sufficiently strong to  determine the
function up to constants. Details of this result can be found in
references [18],[23] and [25].
\par
Using similar techniques one can show very similar results for the
$N=2$ matter. In particular, if we take a restricted form as discussed above
for the form of $H$ one finds that the only four point superconformal
invariants are
 $\hat u$ and $ \hat v$. However, if we take the most general
form of $H$ then
one finds one more independent invariants [18].  Writing
the four point Green's
function as the product of two point functions times an invariant
one can then
show that for a suitable class of low dimension operators that the
function of the invariants is severly restricted [18,23] and is
determined for certain operators [25].
\par
To summarise, we have  formulated the superconformal
Ward identities in the analytic sectors and shown that they and analyticity
place strong constraints on the form that
Green's functions can take.
\par
In this paper we have only discussed superconformal theories.
However, it should be possible  to apply similar
techniques to the spontaneously broken theories and perhaps to the pure
$N=2$ Yang-Mills theory where one has anomalous as well as spontaneous
breaking of superconformal invariance. These topics are under
investigation [24] and it is hoped to  make contact with the work of Seiberg
and Witten [6,7]. It is possible that the non-perturbative methods outlined in
this paper could shed further light on duality, which could be seen as an
inevitable consequence of the powerful constraints resulting from the
symmetries of the theories.
\medskip
{\bf {Note Added}}
Although all the calculations given in the original paper were correct
some conjectures and claims for future work were not. In the revised version we 
have changed these statements so
that they are, with the benefit of hindsight, correct and we have
referred to  where these results can be found in our later work.
In fact, it turns out that extremal Green's functions are soluble as a result of
analyticity and superconformal Ward identities [25] so that the
general thrust of the conjecture in the original paper is correct for some 
special choices of the charges of the operators. These correlators were 
discussed from the AdS point of view in [26,27] and a perturbative verification 
at the one loop level was given in [28].
However, in the original paper, the conjecture was claimed for a larger class of 
correlation functions. In a recent paper [23] the full details of the 
restrictions that analyticity and superconformal symmetry place on four-point 
functions of charge two operators in $N=2$ have been presented; it is shown 
that, for this choice of charges, these retrictions do not fully determine the 
Green's functions.
\medskip
{\bf {References}}
\medskip
\parskip 0pt
\item {[1]} F. Gliozzi, J. Scherk, and D. Olive, Nucl. Phys
B122,(1911), 253; L. Brink, J. Schwarz and J. Scherk, Nucl. Phys
B121, (1977) 77.
\item {[2]} A. Salam and J. Strathdee, Phys. Lett. 51B (1974) 353;
P. Fayet, Nucl. Phys. B113 (1976) 135.
\item {[3]} P. Fayet, Nucl Phys. B113 (1976) 135.
\item {[4]} M. Sohnius and P. West Phys. Lett. B100 (1981) 45; S.Mandelstam,
Nucl. Phys. B213 (1983) 149; P.S. Howe, K.S. Stelle
and P.K. Townsend, Nucl. Phys. B214 (1983) 519; Nucl. Phys. B236
(1984) 125. L. Brink, O. Lindgren and B. Nilsson, Nucl. Phys. B212
(1983) 401.
\item {[5]} P. Howe. K. Stelle and P. West, Phys. Lett 124B (1983) 55.
\item {[6]} N. Seiberg and E. Witten, Nucl. Phys. B246 (1994) 426.
\item {[7]} N. Seiberg and E. Witten, Nucl. Phys. B431 (1994) 484.
\item {[8]} A. Polyakov, J.E.T.P Lett. 12 (1970) 381.
\item {[9]} A. Belavin., A. Polyakov and A. Zamalodchikov, Nucl.Phys.
B241 (1984) 333.
\item {[10]} P. Howe and P. West, Phys. Lett.  B223 (1989) 377.
\item {[11]} B. Conlong and P. West, J. Phys. A26 (1993) 3325.
\item {[12]} S. Ferrara, J.Iliopoulos and B. Zumino, Nucl. Phys. B77
(1974) 413.
\item {[13]} A. Galperin, E. Ivanov, S. Kalitzin, V. Ogievetsky
and E. Sokatchev, Class. Quant. Grav. 2 (1985) 155, Class. Quant. Grav.1 (1984)
469.
\item {[14]} G. Hartwell and P. Howe I.J.M.P to be published.
\item {[15]} J. Wess and B. Zumino, Phys. Lett.  B49 (1974) 52;
J.Iliopoulos and B. Zumino, Nucl. Phys. B76 (1974) 310;
P. West  Nucl. Phys. B106 (1976) 219;
M. Grisaru, M. Ro\v cek and W. Siegel  Nucl. Phys. B159 (1979) 429;
\item {[16]} P. West Phys. Lett. B258 (1991) 369;
I. Jack, T. Jones and P. West, Phys. Lett. B258 (1991) 375;
P. West, Phys. Lett. B261 (1991) 396;
L. Dixon V. Kaplunovsky and J. Louis, Nucl. Phys. B355 (1991) 649;
M. Shifman and A. Vainshtein, Nucl. Phys. B359 (1991) 571.
\item {[17]} P. Howe and G. Hartwell, Class. Quant. Grav. 12
(1995) 1823.
\item {[18]} P. Howe and P. West, Phys.Lett. B400 (1997) 307-313. 
\item {[19]} H. Osborn, Ann. Phys. 231 (1994) 311 and references therein.
\item {[20]} M.T. Grisaru and W.Siegel, Nucl. Phys. B201 (1982) 292.
\item {[21]} P.S. Howe, K.S. Stelle and P.K. Townsend, Nucl. Phys. B214 (1983)
519; Nucl. Phys. B236 (1984) 125.
\item {[22]} P. West ``Supersymmety and Finiteness'' Proceedings of the 1983
Shelter Island II Conference on Quntum Field Theory and Fundamental
Problems of Physics, edited by R. Jackiw, N. Kuri , S. Weinberg and
E. Witten (M.I.T. Press).
\item {[23]} B.U. Eden, P.S.  Howe, A. Pickering, E. Sokatchev and P. West, 
``Four-point functions in $N=2$ superconformal field theories'', hep-th/0001138.
\item {[24]} P. Howe and P. West, ``Superconformal Ward Identities and $N=2$
Yang-Mills Theory'', hep-th/9607239.
\item {[25]}B. Eden, P.S. Howe, C. Schubert,
E. Sokatchev, P. West, ``Extremal correlators in four-dimensional SCFT'', 
hep-th/9910150
\item{[26]}
E. D'Hoker, D.Z. Freedman, S. D. Mathur, A. Matusis and L. 
Rastelli, ``Extremal correlators in the AdS/CFT
correspondence'', MIT-CTP-2893, hep-th/9908160.
\item{[27]}
G. Arutyunov and S. Frolov, ''Some cubic couplings in type IIB
supergravity on $AdS_5 \times S^5$ and three-point functions in $SYM_4$
at large $N$'', LMU-TPW 99-12, hep-th/9907085; ``Scalar quartic
couplings in type IIB supergravity on $AdS_5 \times S^5$'', LMU-TPW 99-23,
hep-th/9912210.
\item{[28]}
M. Bianchi and S. Kovacs,``Nonrenormalization of extremal 
correlators in $N=4$ SYM theory'', ROM2F-99-31, hep-th/9910016. end

\end